# Cross Sections for Nuclide Production in 1 GeV Proton-Irradiated $^{208}$Pb and 0.8 GeV Proton-Irradiated $^{197}$Au

Yu.E. Titarenko, V.F. Batyaev, V.M. Zhivun, A.B. Koldobsky, Yu.V. Trebukhovsky, E.I. Karpikhin, R.D. Mulambetov, S.V. Mulambetova, Yu.V. Nekrasov, A.Yu Titarenko, K.A. Lipatov[1], S.G. Mashnik[3], R.E. Prael [3], K. Gudima[2], M. Baznat [2],

[1] Institute for Theoretical and Experimental Physics (ITEP), B.Cheremushkinskaya 25, 117259 Moscow, Russia
[2] Institute of Applied Physics Academy of Science of Moldova, Chisinau, Moldova
[3] Los Alamos National Laboratory, Los Alamos, NM 87545, USA

***Abstract -****The results of experimental determining the cross sections for residual nuclide production in 1.0 GeV proton-irradiated $^{208}$Pb and 0.8 GeV proton-irradiated $^{197}$Au are presented. The reaction products were recorded by direct γ-spectrometry using a Ge detector of 1.8 keV resolution in the 1332 keV $^{60}$Co γ-line. The experimental results are compared with the experimental data of GSI (inverse kinematics, interactions of accelerated 1.0 GeV/A $^{208}$Pb and 0.8 GeV/A $^{197}$Au ions with a liquid hydrogen target and subsequent recording of of the mass and charges of the emitted product nuclei) and ZSR (direct kinematics, 1.0 GeV and 0.8 GeV proton interactions with $^{nat}$Pb and $^{197}$Au and subsequent γ-spectrometry of the irradiated samples). All the experimental datasets are compared among each other and with the CEM2k, LAQGSM, and LAHET code simulation results.*

## INTRODUCTION

Development and improvement of high-energy transport codes is only possible when supported by reliable experimental data to be used to verify the codes. The cross sections for residual nuclide production measured at different scientific centers (Gesellschaft fuer Schwerionenforschung, Darmstadt, GSI;, Zentrum fuer Strahlenschutz und Radiooekologie Universitaet Hannover, ZSR; and Institute for Theoretical and Experimental Physics, Moscow, ITEP) are most frequently used as such data.

The ITEP and ZSR measurements of the production cross sections of residual product nuclei are based on the analytical method that involves proton irradiation of experimental samples and subsequent γ-spectrometric analysis (called henceforth the direct kinematics method) [1-4]. The GSI approach (the inverse kinematics method) is fundamentally different and involves accelerated ion irradiation of a liquid hydrogen target with subsequent recording of the mass and charges of the target-emitted residual product nuclei without moderating them by physical techniques [5,6].

It is topical in this connection to compare all the experimental datasets among each other and with the CEM2k, LAQGSM, and LAHET code simulation results because the comparison data affect directly the predictive power factor estimates

## COMPARISON BETWEEN THE «DIRECT AND INVERSE KINEMATICS» METHODS

The features of the direct and inverse kinematics methods are described in detail in [1-6]. It should be noted that the direct kinematics method makes it possible to determine mainly the cumulative cross sections and a small fraction of independent cross sections, whereas the inverse kinematics method permits determination of independent cross sections only. The relation

$$\sigma_n^{cum} = \sigma_n^{ind} + \sum_{i=1}^{n-1} \sigma_i^{ind} \prod_{j=i}^{n-1} \nu_j \qquad (1)$$

holds between the independent and cumulative cross section. Here, $\sigma_n^{cum}$ and $\sigma_n^{ind}$ are cumulative and independent cross sections, respectively; $i$ indicates a member of a chain of $(n-1)$ precursors of the given $n$-th nuclide; $\nu_j$ is the branching ratio for decay of the $j$-th member of the chain (j<n).

When calculating the errors in the cumulative cross sections obtained by Formula (1), the independent cross section errors were divided into the statistical errors (from ~ 3% to 25% for $^{208}$Pb and $^{197}$Au) and the total error in the correction factors (~ 9% at Z = 82-61 for $^{208}$Pb and at Z = 79-60 for $^{197}$Au; ~15% at Z values characteristic of fission products; ~25% at Z values intermediate between the Z= 61, 60 and the fission range of Z). The statistical errors were calculated by the standard formulas. The correction factor error was allowed for during the last stage of calculating the total error.

Since the inverse kinematics method fails to separate the metastable and ground states, some data were excluded from the comparison,

The experimental $^{208}Pb$ and $^{197}Au$ data obtained by different Laboratories are presented, respectively, in [2] and in Table 1 below.

Table 1. The ITEP experimental production cross sections (mbarn) of the $^{197}Au(p,x)$ reactions at $E_p$ = 0.8 GeV as compared with the GSI and ZSR results

| Product | $T_{1/2}$ | Type | ITEP | GSI | ZSR |
|---|---|---|---|---|---|
| $^{195m}Hg$ | 41.6 h | i(m) | 4.46±0.50 | - | - |
| $^{193m}Hg$ | 11.8 h | i(m) | 5.86±0.70 | - | - |
| $^{192}Hg$ | 4.85 h | i | 6.92±0.51 | 4.22 | - |
| $^{198}Au$ | 2.69517 d | i(m+g) | 3.64±0.23 | - | - |
| $^{196}Au$ | 6.183 d | i(m1+m2+g) | 82.8±5.0 | 58.0 | 66.0±7.0 |
| $^{195}Au$ | 186.10 d | c | 58.6±5.2 | 41.1 | 52.5±6.1 |
| $^{194}Au$ | 38.02 h | i(m1+m2+g) | 40.6±2.7 | 31.0 | 36.2±6.2 |
| $^{193}Au$ | 17.65 h | i(m+g) | 22.5±3.0 | 29.4 | - |
| $^{193}Au$ | 17.65 h | c | 27.0±3.3 | 33.8 | - |
| $^{192}Au$ | 4.94 h | i(m+g) | 32.7±3.1 | 25.2 | - |
| $^{192}Au$ | 4.94 h | c | 39.4±3.7 | 29.4 | - |
| $^{191}Au$ | 3.18 h | c | 32.5±2.2 | 27.5 | - |
| $^{190}Au$ | 42.8 min | c | 26.3±2.5 | 25.3 | - |
| $^{191}Pt$ | 2.802 d | c | 45.0±3.7 | 47.4 | 31.1±4.2 |
| $^{189}Pt$ | 10.87h | c | 57.7±4.7 | 43.4 | - |
| $^{188}Pt$ | 10.2d | c | 48.0±3.0 | 39.7 | 46.0±3.7 |
| $^{186}Pt$ | 2.08h | c | 30.4±3.3 | 34.7 | - |
| $^{192}Ir$ | 73.827d | i(m1+m2+g) | 2.76±0.17 | 4.18 | 2.91±0.23 |
| $^{190}Ir$ | 11.78d | i(m1+m2+g) | 4.36±0.27 | 6.02 | - |
| $^{189}Ir$ | 13.2d | c | 50.6±4.8 | 51.5 | 58.7±4.7 |
| $^{188}Ir$ | 41.5h | i | 11.8±0.8 | 8.97 | - |
| $^{188}Ir$ | 41.5h | c | 62.0±4.0 | 48.6 | - |
| $^{186}Ir$ | 16.64h | c | 28.4±1.8 | - | - |
| $^{186m}Ir$ | 1.90h | c | 30.4±5.2 | - | - |
| $^{185}Ir$ | 14.4h | c* | 41.4±3.8 | 47.5 | - |
| $^{184}Ir$ | 3.09h | c* | 49.2±3.3 | 46.7 | - |
| $^{185}Os$ | 93.6d | c | 60.5±3.8 | 49.8 | 56.9±5.6 |
| $^{183m}Os$ | 9.9h | i(m) | 33.2±2.1 | - | - |
| $^{183}Os$ | 13.0h | i(m+g) | 25.2±1.7 | - | - |
| $^{183}Os$ | 13.0h | c | 32.6±2.0 | - | - |
| $^{182}Os$ | 22.10h | c | 58.4±3.7 | 50.1 | - |
| $^{181}Os$ | 105m | c | 16.4±1.8 | - | - |
| $^{183}Re$ | 70.0d | c | 62.9±4.1 | 52.0 | 49.7±4.3 |
| $^{182m}Re$ | 12.7h | c | 61.5±4.1 | - | - |
| $^{182}Re$ | 64.0h | i | 6.35±0.68 | - | - |
| $^{181}Re$ | 19.9h | c | 56.9±5.5 | 51.3 | - |
| $^{179}Re$ | 19.5m | c* | 68.8±5.8 | 69.7 | - |
| $^{178}W$ | 21.6d | c | 43.2±3.6 | 49.4 | - |
| $^{177}W$ | 135m | c* | 47.3±5.4 | 39.7 | - |
| $^{177}Ta$ | 56.56h | c | 57.0±10.0 | 36.8 | - |
| $^{176}Ta$ | 8.09h | c | 55.1±4.0 | 46.3 | - |
| $^{175}Ta$ | 10.5h | c | 54.7±3.9 | 42.2 | - |
| $^{174}Ta$ | 1.14h | c | 52.4±4.7 | 41.9 | - |
| $^{173}Ta$ | 3.14h | c* | 59.5±4.2 | 43.6 | - |
| $^{172}Ta$ | 36.8m | c* | 29.2±2.8 | 41.6 | - |
| $^{175}Hf$ | 70d | c | 52.9±3.5 | 45.4 | 44.7±4.5 |
| $^{173}Hf$ | 23.6h | c | 50.0±3.2 | 41.7 | - |
| $^{172}$Hf | 1.87y | c | 44.6±2.8 | 40.1 | 33.9±2.9 |
| $^{170}$Hf | 16.01h | c | 40.6±2.9 | 35.4 | - |
| $^{173}$Lu | 1.37y | c | 47.6±3.1 | 41.8 | 68.8±6.8 |
| $^{172}$Lu | 6.70d | c | 45.2±2.9 | 40.3 | - |
| $^{171}$Lu | 8.24d | c* | 48.1±2.9 | 29.6 | 43.9±4.2 |
| $^{170}$Lu | 2.012d | c | 40.0±2.6 | 34.4 | - |
| $^{169}$Lu | 34.06h | c | 34.0±2.1 | 20.7 | - |
| $^{167}$Lu | 51.5m | c | 36.7±2.7 | 29.7 | - |
| $^{169}$Yb | 32.026d | c | 39.4±2.4 | 31.8 | 34.6±3.8 |
| $^{166}$Yb | 56.7h | c | 32.0±2.0 | 26.0 | 19.9±1.4 |
| $^{162}$Yb | 18.87m | c | 25.0±3.1 | 16.9 | |
| $^{168}$Tm | 93.1d | i | - | 0.04 | 0.283±0.047 |
| $^{167}$Tm | 9.25d | c | 40.2±4.0 | 26.2 | 30.0±3.2 |
| $^{166}$Tm | 7.70h | c | 34.1±2.1 | 28.3 | - |
| $^{165}$Tm | 30.06h | c | 30.2±2.0 | 25.6 | - |
| $^{161}$Er | 3.21h | c* | 25.8±2.1 | 19.5 | - |
| $^{160}$Er | 28.58h | c | 19.7±1.6 | 14.3 | - |
| $^{159}$Er | 36m | c* | 19.8±1.9 | 15.8 | - |
| $^{160m}$Ho | 5.02h | c | 19.6±1.8 | - | - |
| $^{156}$Ho | 56m | c | 11.2±0.9 | 8.03 | - |
| $^{157}$Dy | 8.14h | c | 12.0±0.8 | 9.53 | - |
| $^{155}$Dy | 9.9h | c* | 8.68±0.56 | 5.38 | - |
| $^{152}$Dy | 2.38h | c | 4.24±0.36 | - | - |
| $^{155}$Tb | 5.32d | c | 7.80±0.58 | 4.98 | - |
| $^{153}$Tb | 2.34d | c* | 5.52±0.42 | 4.31 | 3.27±0.41 |
| $^{152}$Tb | 17.5h | c | 3.80±0.38 | - | - |
| $^{151}$Tb | 17.609h | c | 1.85±0.19 | - | - |
| $^{153}$Gd | 240.4d | c | 4.94±0.39 | 3.84 | - |
| $^{149}$Gd | 9.28d | c | 3.91±0.30 | - | 4.49±0.34 |
| $^{146}$Gd | 48.27d | c | 2.63±0.17 | 1.84 | 2.13±0.19 |
| $^{149}$Eu | 93.1d | c | - | - | 7.22±0.62 |
| $^{148}$Eu | 55.6d | i | - | 0.08 | 0.149±0.014 |
| $^{147}$Eu | 24.1d | c* | 3.08±0.30 | 1.61 | 2.66±0.25 |
| $^{146}$Eu | 4.61d | i | 0.297±0.054 | 0.210 | - |
| $^{146}$Eu | 4.61d | c | 2.83±0.18 | 1.62 | - |
| $^{145}$Eu | 5.93d | c | 1.69±0.16 | 1.72 | 1.28±0.12 |
| $^{143}$Pm | 265d | c | - | 1.24 | 0.953±0.079 |
| $^{139}$Ce | 137.640d | c | 0.757±0.052 | 0.650 | 0.569±0.043 |
| $^{127}$Xe | 36.4d | c | - | 0.07 | 0.887±0.084 |
| $^{121}$Te | 16.8d | c | 0.714±0.087 | 0.204* | 0.430±0.030 |
| $^{121m}$Te | 154d | i(m) | - | - | 0.212±0.019 |
| $^{113}$Sn | 115.1d | c | - | 0.184* | 1.38±0.45 |
| $^{111}$In | 2.8047d | c | 0.545±0.049 | 0.265 | - |
| $^{110m}$Ag | 249.79d | i | - | - | 0.384±0.056 |
| $^{105}$Ag | 41.29d | c | 1.14±0.11 | 0.29 | 1.70±0.23 |
| $^{101m}$Rh | 4.34d | c | 0.851±0.108 | - | 1.36±0.15 |
| $^{103}$Ru | 39.26d | c | 1.22±0.12 | 0.589 | 1.22±0.10 |
| $^{96}$Tc | 4.28d | i(m+g) | 0.848±0.063 | 0.584* | 0.784±0.063 |
| $^{95}$Nb | 34.997d | i(m+g) | 1.34±0.11 | 0.823* | - |
| $^{95}$Nb | 34.997d | c | 2.03±0.13 | 1.12* | 1.53±0.21 |
| $^{95}$Zr | 64.02d | c | 0.686±0.067 | 0.292 | 0.655±0.149 |
| $^{89}$Zr | 78.41h | c | 1.67±0.10 | 1.27* | 1.63±0.13 |
| $^{88}$Zr | 83.4d | c | 0.880±0.103 | 0.725 | 0.724±0.073 |
| $^{88}$Y | 106.65d | i | 1.98±0.14 | 1.39 | 2.45±0.19 |
| $^{88}$Y | 106.65d | c | 2.82±0.20 | 2.12 | - |
| $^{87}$Y | 79.8h | c* | 2.38±0.15 | 1.27 | 1.72±0.20 |
| $^{85}$Sr | 64.9d | c | 2.34±0.16 | 1.32* | 1.85±0.14 |
| $^{86}$Rb | 18.631d | i(m+g) | - | 0.99 | 3.04±0.53 |
| $^{83}$Rb | 86.2d | c | 2.67±0.24 | 1.81 | 2.01±0.14 |
| $^{82m}$Rb | 6,472h | i(m) | 1.18±0.13 | - | - |
| $^{82}$Br | 35.30h | i(m+g) | 0.758±0.056 | 0.782* | - |
| $^{75}$Se | 119.779d | c | 1.17±0.08 | 0.909 | 0.970±0.070 |
| $^{74}$As | 17.77d | i | 1.42±0.11 | 1.10 | 1.37±0.11 |
| $^{65}$Zn | 244.3d | c | - | 0.47 | 0.554±0.160 |
| $^{60}$Co | 5.27y | i(m+g) | - | 0.533 | 0.749±0.089 |
| $^{58}$Co | 70.8d | i(m+g) | - | 0.253 | 0.960±0.095 |
| $^{59}$Fe | 44.5d | c | - | 0.48 | 0.503±0.032 |
| $^{54}$Mn | 312d | i | - | 0.312 | 0.390±0.045 |
| $^{46}$Sc | 83.8d | i(m+g) | - | 0.255 | 0.166±0.016 |
| $^{22}$Na | 2.60y | c | - | - | 0.050±0.005 |

*The value have been obtained assuming a small (<15%) ratio of branching into metastable levels.

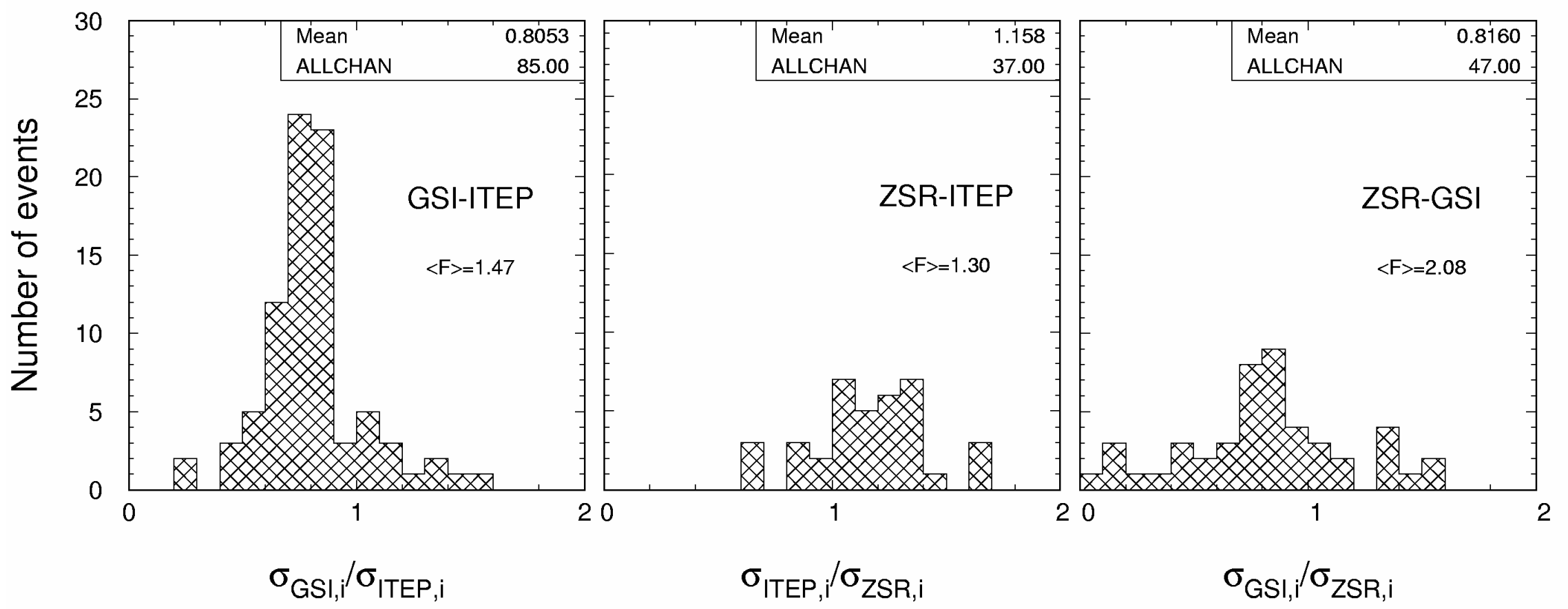

Fig. 1. Results of comparing among the experimental cross sections for residual nuclide production in $^{197}$Au(p,x) reaction at $E_p$ = 0.8 GeV (ITEP, $^{208}$Pb; ZSR, $^{nat}$Pb) and in $^{1}$H($^{197}$Au,x) reaction at $E_{Au}$=0.8 GeV/A (GSI)

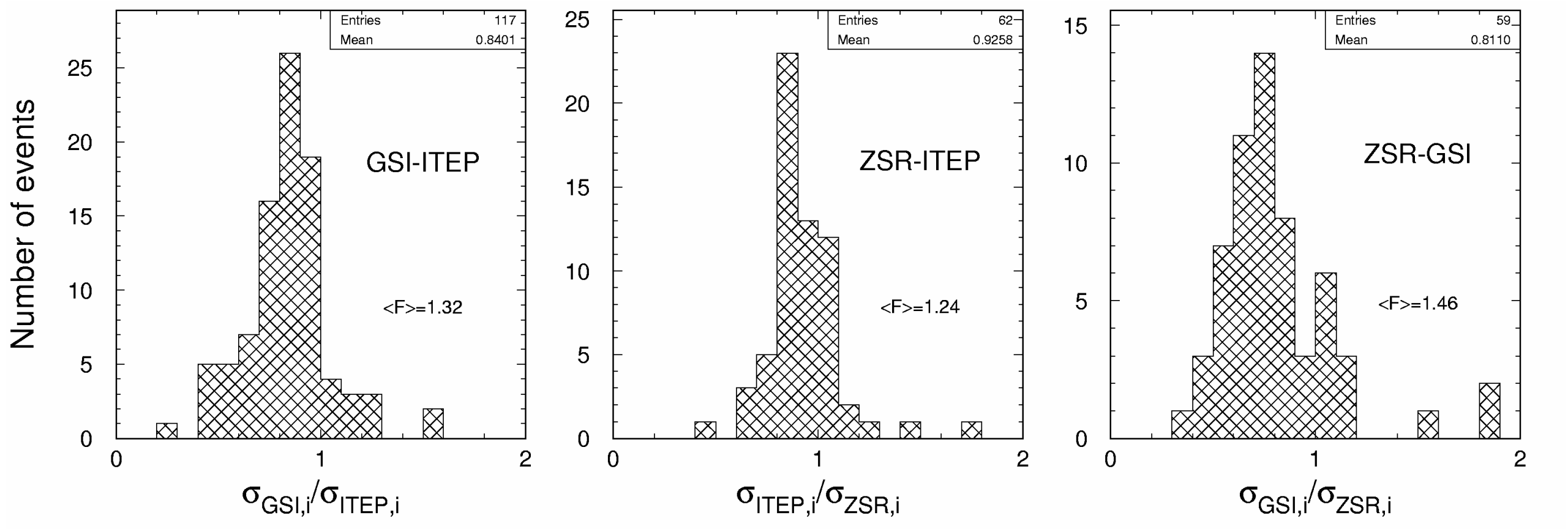

Fig. 2. Results of comparing among the experimental cross sections for residual nuclide production in $^{208}$Pb(p,x) reaction at $E_p$ = 1.0 GeV (ITEP, $^{208}$Pb; ZSR, $^{nat}$Pb) and in $^{1}$H($^{208}$Pb,x) reaction at $E_{Au}$= 1.0 GeV/A (GSI)

An important requirement of any correct comparison between the results of the above two methods is that the projectile proton energy of the direct kinematics method should be about the same as the energy per a nucleon of accelerated ion interactions with the liquid H target of the inverse kinematics method. It can readily be seen that this requirement is met in the described experiments.

Besides, it is desirable that the nucleonic composition of accelerated ions of the inverse kinematics method should correspond to the isotopic composition of the proton-irradiated targets of the direct kinematics method. In the case of $^{197}$Au, this condition is satisfied in all the analyzed works because Au is monoisotopic. In the case of $^{208}$Pb, the condition is fully satisfied only in the GSI-ITEP comparison because the ITEP target is made of lead high-enriched with $^{208}$Pb (97.2%). In the ZSR experiments, the condition is satisfied up to the $^{208}$Pb content (52.4%) of natural lead.

The results of comparing the experimental cross sections for residual product nuclide production are shown in Fig. 1 for the $^{197}$Au(p,x) reaction at $E_p$ = 0.8 GeV (ITEP, ZSR) and for $^1$H($^{197}$Au,x) at $E_{Au}$ = 0.8 GeV/A (GSI) and in Fig. 2 for the $^{208}$Pb(p,x) reaction at $E_p$ = 1.0 GeV (ITEP,$^{208}$Pb; ZSR,$^{nat}$Pb ) and for the $^1$H($^{208}$Pb,x) reaction at $E_{Pb}$ = 1.0 GeV/A (GSI).

The results are presented as plots of the statistics of the ratios of residual nuclide production cross sections for each pair of the compared datasets. The values of the mean squared deviation factor <F> in the histograms were calculated by the method proposed in [4].

## SIMULATION OF EXPERIMENTAL DATA

The results of measuring the production cross sections of secondary product nuclei were simulated by the INCL model of the LAHET [7], CEM2k [8], and LAQGSM [9] high-energy codes that are used extensively at intermediate and high energies to simulate the intermediate- and high-energy hadron-nucleus interactions. Figs. 3-8 show the experiment-simulation comparison results.

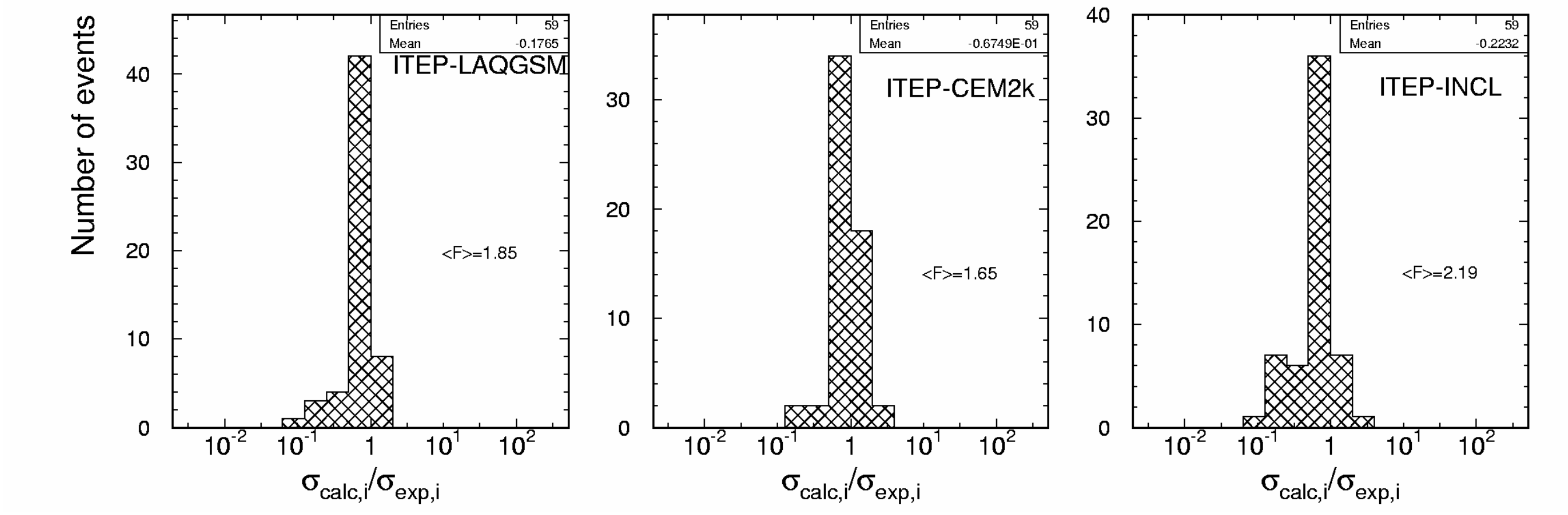

Fig. 3. Results of comparing the experimental cross sections for residual nuclide production in $^{197}$Au(p,x) reaction at $E_p$ = 0.8 GeV (ITEP) with the LAQGSM, CEM2k, and LAHET code simulation results.

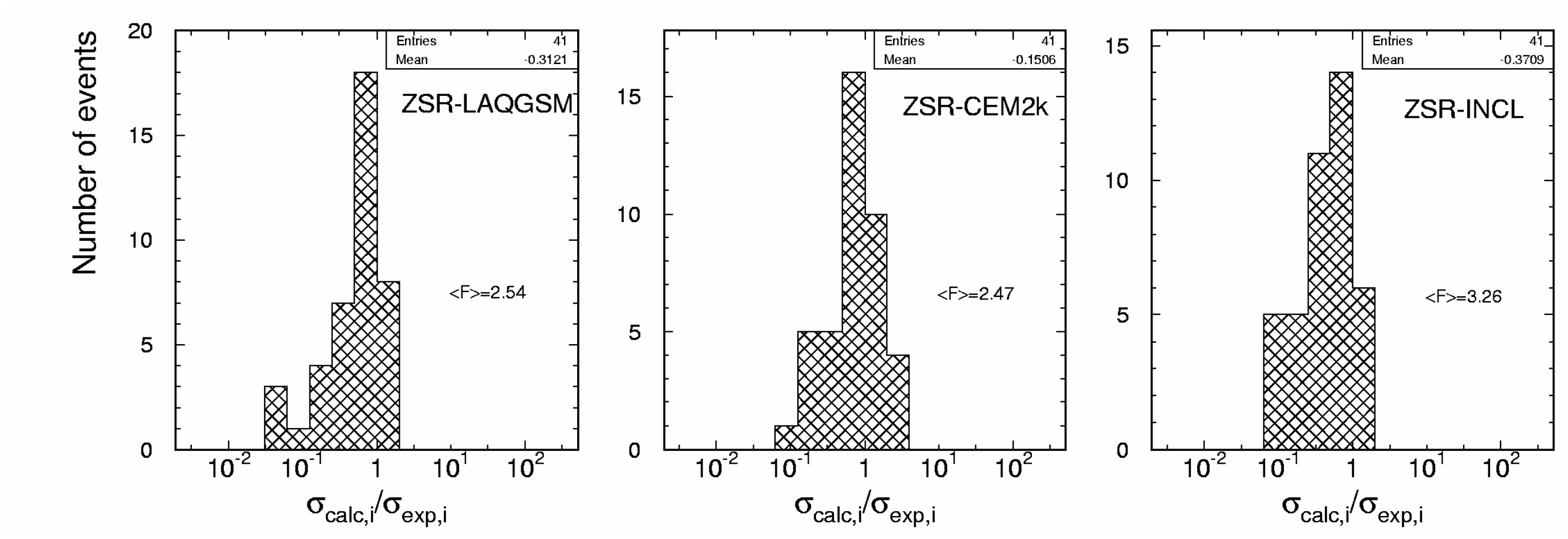

Fig. 4. Results of comparing the experimental cross sections for residual nuclide production in $^{197}$Au(p,x) reaction at $E_p$ = 0.8 GeV (ZSR) with the LAQGSM, CEM2k, and LAHET code simulation results.

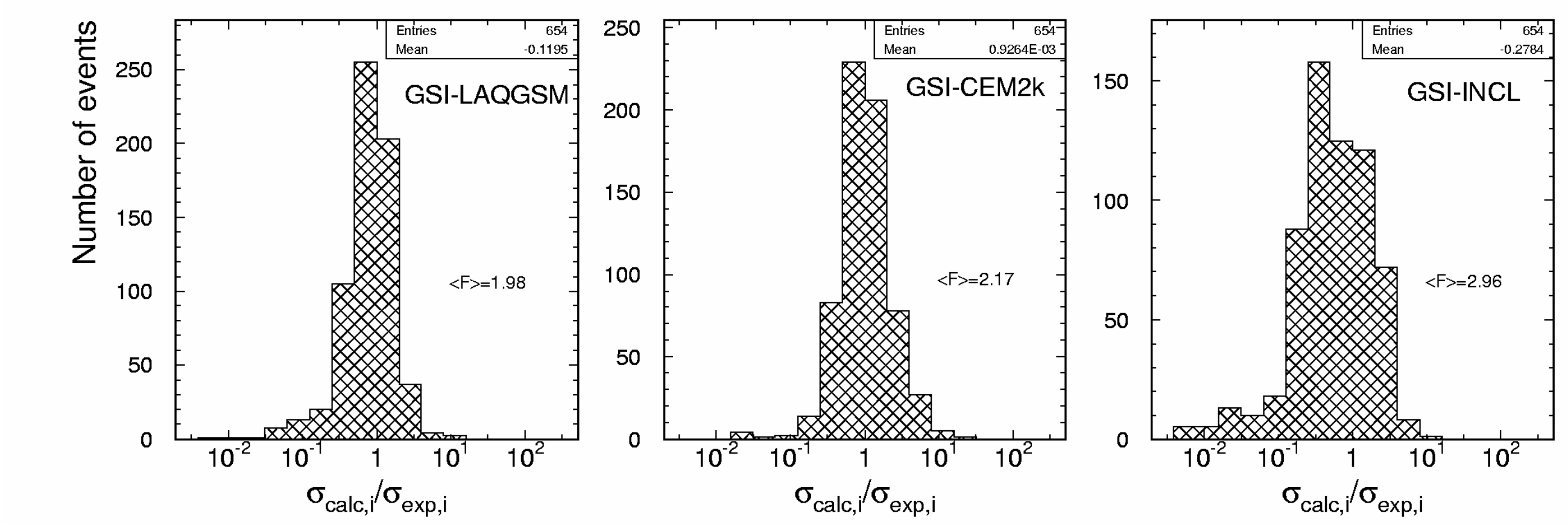

Fig. 5. Results of comparing the experimental cross sections for residual nuclide production in $^{197}$Au(p,x) reaction at $E_p$ = 0.8 GeV (ITEP) with the LAQGSM, CEM2k, and LAHET code simulation results.

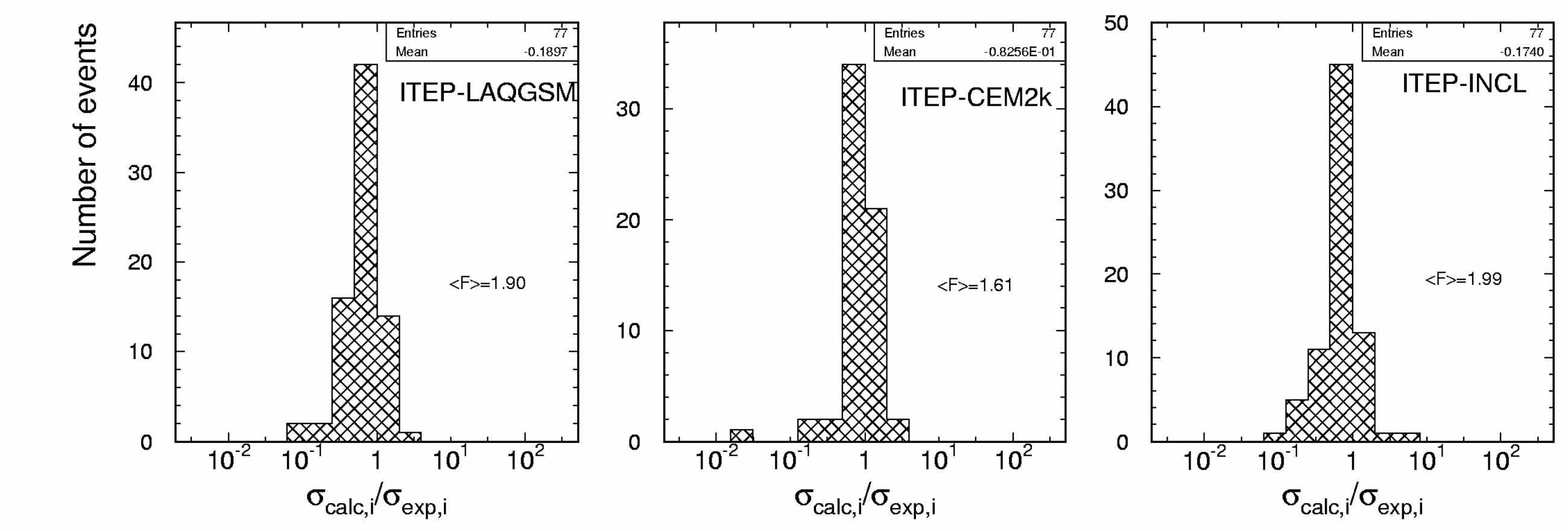

Fig. 6. Results of comparing the experimental cross sections for residual nuclide production in $^{208}$Pb(p,x) reaction at $E_p$ = 1.0 GeV (ITEP) with the LAQGSM, CEM2k, and LAHET code simulation results.

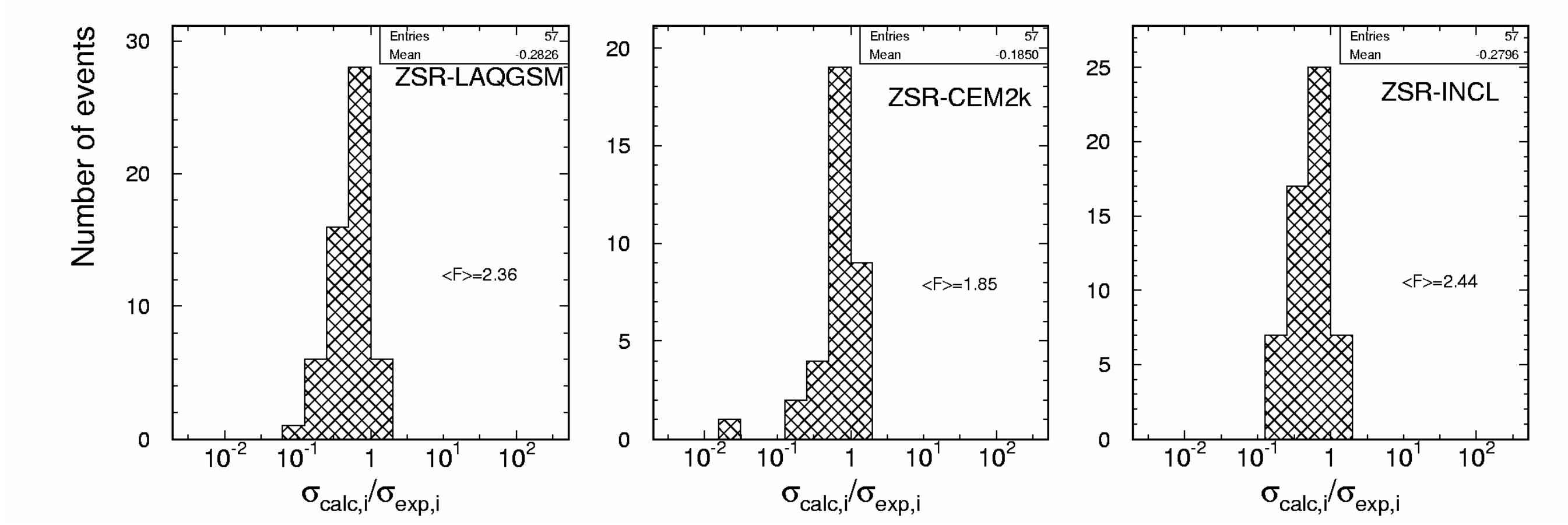

Fig. 7. Results of comparing the experimental cross sections for residual nuclide production in $^{208}Pb(p,x)$ reaction at $E_p$ = 1.0 GeV (ZSR) with the LAQGSM, CEM2k, and LAHET code simulation results.

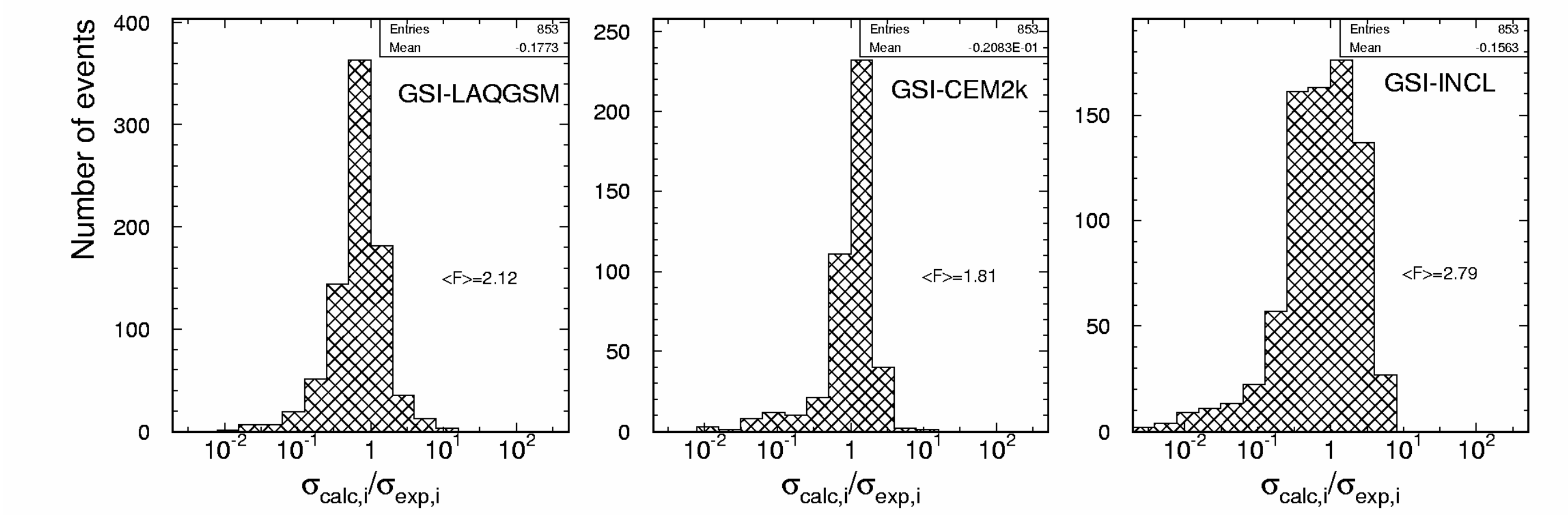

Fig. 8. Results of comparing the experimental cross sections for residual nuclide production in $^{1}H(^{208}Pb,x)$ reaction at $E_{Pb}$ = 1.0 GeV/A (GSI) with the LAQGSM, CEM2k, and LAHET code simulation results.

## CONCLUSIONS

The comparative analysis of the experimental results of determining the residual product nuclide production cross sections by the above discussed two methods (Figs. 1 and 2) permits the following qualitative conclusions.

- The general agreement among the experimental data of three laboratories may be considered satisfactory, with the residual product nuclide production cross sections obtained by the direct kinematics method being in the average ~20% higher that those obtained by the inverse kinematics method. This is evident because the comparison between the ITEP and ZSR results obtained both by the direct kinematics method fails to show any systematic differences (the ITEP data are higher for Au, and vice versa for Pb). The noted differences have probably arisen from the systematic errors inherent just to the direct kinematics method. At the same time, it seems impossible now to estimate the errors, thus invoking additional studies.
- Comparison of the simulation results of the tested codes with the experimental data of three laboratories (Figs. 3-8) has shown that the agreement is in all cases much worse than in the case of comparison among the experimental data. Obviously, the model representations of hadron-nucleus interactions that underlay the codes are still subject to substantial modification (this conclusion is additionally substantiated by the high values of mean squared deviation factor <F> in the calculation-experiment comparisons (~2.2 in the average) as confronted to the averaged <F>value (~1.5) in the comparison among the experimental data of the three laboratories).
- The fact attracts attention that, when compared, the ZSR experimental Au data show not only signifi-

cant systematic differences from both ITEP and GSI, but also (in the comparison with GSI) a high mean squared deviation factor <F>. The ZSR $^{197}$Au data at $E_p$ = 0.8 GeV may be assumed to suffer definite systematic errors that lead to underestimated cross sections. This can be indirectly confirmed, for instance, by analyzing the data of [4].

This work was partially supported by the U.S. Department of Energy and by the Moldovan-U.S. Bilateral Grants Program, CRDF Project MP2-3025.